\documentclass[11pt]{article}

\usepackage{aas_macros}
\usepackage{lineno}
\usepackage{natbib}
\usepackage{soul}
\usepackage{hhline}
\usepackage{url}
\usepackage{amsmath,amssymb}
\usepackage{bm}
\usepackage{subcaption}
\usepackage{comment}
\usepackage{graphicx}
\usepackage[dvipsnames]{xcolor}
\usepackage{hyperref}
\usepackage[left=2cm, right=2cm, top=2.5cm, bottom=2.5cm]{geometry}
\newcommand{\kmps}{\text{km/s}}

\begin{document}

%%%%%%%%%%%%%%%%%%% TITLE PAGE %%%%%%%%%%%%%%%%%%%

% Title of the paper, and the short title which is used in the headers.
% Keep the title short and informative.
\title{A robust empirical relationship between speed and turbulence energy in the near-Earth solar wind}

% The list of authors, and the short list which is used in the headers.
% If you need two or more lines of authors, add an extra line using \newauthor
\author{Rohit Chhiber$^{1,2}$\thanks{\url{rohit.chhiber@nasa.gov},\url{rohitc@udel.edu}}~~, Yanwen Wang$^{3}$,
Manuel E. Cuesta$^{4}$,
Jiaming Wang$^{1}$,
and Sohom Roy$^{5}$
\\ \\
% List of institutions
$^{1}$Department of Physics and Astronomy, University of Delaware, Newark, DE 19716, USA\\
$^{2}$Heliophysics Science Division, NASA Goddard Space Flight Center, Greenbelt, MD 20771, USA\\
$^{3}$Department of Physics, University of Maryland, College Park, MD 20742, USA\\
$^{4}$Department of Astrophysical Sciences, Princeton University, Princeton, NJ 08544, USA \\
$^{5}$Space Research Institute, Austrian Academy of Sciences, Schmiedlstraße 6, 8042, Graz, Austria
}
% These dates will be filled out by the publisher
% Enter the current year, for the copyright statements etc.

% Don't change these lines
%\begin{document}
%\label{firstpage}
\maketitle

\begin{abstract}
The connection between turbulence and solar-wind acceleration, long known in space physics, is further developed in this paper by establishing a robust empirical law that relates the bulk-flow speed to the magnetohydrodynamic-scale fluctuation energy in the plasma. The model is based on analysis of twenty-five years of near-Earth observations by NASA's Advanced Composition Explorer. It provides a simple way to estimate turbulence energy from low-resolution speed data -- a practical approach that may be of utility when high-resolution measurements or advanced turbulence models are unavailable. Potential heliospheric applications include space-weather forecasting operations, remote imaging datasets, and energetic-particle transport models that require turbulence amplitudes to specify diffusion parameters.
\end{abstract}

%\begin{keywords}
%\textit{Keywords}: Sun: photosphere -- Sun: magnetic fields -- turbulence
%\end{keywords}

%%%%%%%%%%%%%%%%% BODY OF PAPER %%%%%%%%%%%%%%%%%%

\section{Introduction}\label{sec:intro}
%
%The main text should start with an introduction. Except for short  manuscripts (such as comments and replies), the text should be divided  into sections, each with its own heading.

The solar wind is the continual dynamical outflow of plasma from the Sun into interplanetary space. It plays a fundamental role in physical processes occurring throughout the Solar System and is the main driver of the terrestrial impacts of space weather \citep{Pulkkinen2007LRSP}. The solar wind is also the only astrophysical system that can be accessed \textit{in situ}, and as such, it serves as a unique and gigantic plasma laboratory in which scientists can investigate a variety of phenomena of relevance to astrophysics and plasma physics \citep[e.g.,][]{bruno2013LRSP,gurnett2017PlasmaPhys_book}.

The basic physical concepts governing the solar wind have been known since the theoretical considerations of \cite{parker1958apj} and subsequent \textit{in-situ} observations in the 1960s \citep[for a historical review, see, e.g.,][]{obridko2017SoSyR}. Later studies have acknowledged the rich complexity of its structure and dynamics \citep[e.g.,][]{Verscharen2019LRSP}. Of particular note are questions relating to the heating and acceleration of the corona and the solar wind \citep[e.g.,]{leer1982ssr,klimchuk2006SoPh}, and the transport of hazardous solar energetic particles (SEPs) and cosmic rays (CRs) through the heliosphere \citep[e.g.,][]{Engelbrecht2022SSR}. 

In recent decades, the study of waves and turbulence in the heliosphere has emerged as a major focus within the broader effort to address the above questions \citep{matthaeus2011SSR}. The turbulent cascade \citep[e.g.,][]{pope2000book} provides an effective mechanism to account for non-adiabatic heating in weakly-collisional space plasmas \citep[][]{Kiyani2015RSPTA,matthaeus2020ApJ}, and it is generally acknowledged that fast wind from coronal holes requires sustained acceleration over an extended region which can be provided by propagating Alfv\'en-wave-like fluctuations \citep[e.g.,][]{DePontieu2007Sci,Cranmer2009LRSP_coronal_holes,Rivera2024Sci,Usmanov2025ApJ}. Interplanetary turbulence can also account for the observed diffusive scattering of SEPs to distant heliolongitudes \cite[e.g.,][]{laitinen2016AA,chhiber2021ApJ_flrw} and directly induce geomagnetic activity \cite[e.g.,][]{Borovsky2003JGRA,DAmicis2007GRL}, both effects of immediate interest to space weather forecasting operations \citep{bothmer2007SpaceWeather}.

While the importance of turbulence in heliospheric phenomena has been recognized, incorporating turbulence within computational models of the solar wind and SEP transport has been challenging. Resolving the full range of spatio-temporal scales involved in magnetohydrodynamic (MHD) turbulence is computationally intractable for global three-dimensional simulations of the heliosphere \citep{miesch2015SSR194,gombosi2018LRSP}. In response, some global simulations have adopted a type of ``subgrid'' modeling approach that couples the resolved bulk flow with  approximate statistical models of unresolved turbulence \citep[e.g.,][]{matthaeus1999ApJL523,usmanov2000global,vanderholst2014ApJ,Shiota2017ApJ837}. By accounting for the low-frequency ``energy containing'' scales of turbulence \citep[e.g.,][]{zhou2004RMP}, these physics-based models reproduce a variety of observations, including non-adiabatic heating and fast wind speeds \citep[e.g.,][]{chhiber2021ApJ_psp,Usmanov2025ApJ}. However, they are mathematically complex and computationally demanding; importantly, the latter aspect makes them unsuitable for real-time space weather forecasting and for community tools that provide ``runs on request'' \citep[e.g.,  NASA's][]{CCMC_webpage}, which require relatively brief run time. The latter class of models typically employs empirical approaches like the Wang-Sheeley-Arge \citep[WSA;][]{Wang1990ApJ,Arge2000JGR} relationship between magnetic flux-tube expansion and wind speed to produce fast solar wind streams, machine learning methods \citep[e.g.,][]{Upendran2020SpWea}, and/or polytropic equations or ad-hoc heating functions to obtain realistic temperatures \citep[e.g.,][]{Pizzo2011SpacWeat,Sokolov2013ApJ,MacNeice2018SpWea,Samara2024ApJ}. SEP forecasting models often use heuristic, constant turbulence levels \citep[e.g.,][]{Hu2022SciAdv,Whitman2023AdSpR}. 

In this paper we develop an empirical approach that can be used to obtain reasonably accurate estimates of MHD turbulence levels simply from low-resolution speed data. It is based on an empirical law, derived from analysis of 25 years of near-Earth observations by NASA's \textit{Advanced Composition Explorer} (\textit{ACE}) mission, that relates the bulk speed of the solar wind to its average turbulence energy. The approach can be useful in the aforementioned modeling frameworks that lack turbulence and also in extracting turbulence information from low-resolution observations. Potential observational applications include sparse \textit{in-situ} datasets from the outer heliosphere \citep[e.g., \textit{Voyager} and \textit{New Horizons};][]{Elliott2016ApJS,Wrench2025ApJ} and flow-speed maps derived from remote-sensing images \citep[e.g.,][]{DeForest2025arXiv_PUNCH,Attie2025AAS}.

We bear in mind that a positive correlation between wind speed and turbulence levels has been noted in previous work \citep[e.g.,][]{Forsyth1996GRL,Horbury2001JGR,Erdos2005AdSpR,Borovsky2012JGR_vb_fluc,Shi2023ApJ,Usmanov2025ApJ}, and it is believed to be related to magnetic topology at the solar source \citep{Wang1991ApJ,Cranmer2019ARAA}. However, to the best of our knowledge, the present study is the first time an empirical model based on this relationship has been derived and evaluated. We delve deeper into its physical origins in a companion paper.  

 \section{Data}\label{sec:data} 

We employ \textit{in-situ} observations from \textit{ACE}, situated at \(\text{L}_1\) and spanning the period from 1998-02-05 00:00:00 to 2023-12-31 23:59:00 (UTC). The magnetic field was obtained from the MAG instrument \citep{smith1998ace} and plasma data (ion number density and velocity) from the SWEPAM instrument \citep{mccomas1998solar}. The downloaded magnetic data with 1-s resolution were down-sampled to a 1-min cadence, while density and velocity data at 64-s were upsampled using linear interpolation to match the magnetic field's cadence. These data span \(\sim2.5\) solar activity cycles \citep{Usoskin2023LRSP_sol_activity}, including three maxima and two minima (see Fig. \ref{fig:time1}). Plasma measurements are sometimes unavailable; possible reasons for this include contamination by large CME/SEP events \citep{skoug2004extremely}, instrumental aging after 2010 \citep{Skoug2014ACErepointing}, or spacecraft attitude adjustment and calibration periods. The instrumental aging that affects proton density measurements was addressed in 2012 with adjustments to the spacecraft by the \textit{ACE} team \citep[see][]{Skoug2014ACErepointing}. 

In our analysis, missing data points are represented as `\(NaN\)', and the full dataset is partitioned into consecutive non-overlapping 12-hour intervals that begin at 00:00 each day, closed at the start time and open at the end time. The \(NaNs\) mainly affect density data, which, for our analysis, are only required to compute the mean density within an interval (see Sec. \ref{sec:res}). We exclude intervals in which the fraction of \(NaNs\) in density is above 90\%. In the remaining intervals the fraction of \(NaNs\) in the velocity and magnetic field data is below 10\%.

Our choice of time-series cadence and averaging interval duration is motivated by known facts regarding the scales of MHD turbulence in the solar wind at 1 AU \citep[see, e.g.,][]{Kiyani2015RSPTA}. The high-frequency end of the inertial range is \(\sim 5\) seconds, while the low-frequency end is \(\sim 30\) minutes. Our focus is on the energy-containing scales of turbulence, so we wish to compute turbulence parameters that include information from at-least a decade of the low-frequency end of the inertial range, which motivates our time-series resolution of 1 minute. This also ensures that we are at sufficiently large scales compared to kinetic scales (ion-inertial length; \(\sim 1\) second at 1 AU). The typical correlation time of turbulence at 1 AU is around 1 hour; accordingly, the averaging duration of 12 hours is chosen to include several correlation scales. The requirement of including several correlation scales within an averaging interval to obtain well-behaved turbulence statistics is described in detail by, e.g., \cite{matthaeus1982JGR}. Of course, these choices are not rigidly defined. Accordingly, we have repeated our analysis (Sec. \ref{sec:res}) for 6 and 24 hour intervals using 1-min time-series resolution, and changing the temporal resolution to 3 minutes (by first smoothing the 1-min time-series over a 3-min moving window and then down-sampling) while keeping a 12-hour averaging duration. We also repeated the analysis by excluding intervals in which the fraction of \(NaNs\) in density data is above 70\%. The results are not significantly changed in any of these cases.

We identify intervals  containing interplanetary coronal mass ejections \citep[ICMEs;][]{Webb2012LRSP_CMEs} by referencing a catalog compiled by \cite{Richardson2010SoPh}, which has been updated as recently as 2025 \citep{Richardson2024CMElist}; 1,270 such intervals are identified. The remainder of the intervals (10,661 in number) are considered to be representative of the ambient solar wind in near-Earth space, which is our focus here. The resulting \textit{ACE} dataset has been employed in a number of recent studies by our group \citep{roy2021ApJl,Roy2022PRE,Wang2024ApJ,Wang2026PNAS,Wang2026arXiv}.

 \section{Results}\label{sec:res}

To examine the correlation between bulk solar wind speed and turbulence energy we compute the mean speed for each 12-hour interval: \(V\equiv |\bm{V}|\equiv |\langle \tilde{\bm{V}}\rangle|\), where the \(\langle\cdot\rangle\) operator indicates an arithmetic mean computed over an interval and  \(\tilde{\cdot}\) represents a quantity at 1-min cadence (in this case the ion velocity \(\tilde{\bm{V}}\)). The magnetic field \(\tilde{\bm{B}}\) is converted to Alfv\'en units: \(\tilde{\bm{B}}_A=\tilde{\bm{B}}/\sqrt{4\pi\rho}\) where \(\rho=m_pn \equiv  m_p\langle\tilde{n}\rangle\) is the mean proton mass density of the interval, computed from proton mass \(m_p\) and ion number density \(\tilde{n}\). Hereafter, it is understood that the magnetic field is in Alfv\'en units and we freely drop the subscript `\(A\)'. Magnetic and velocity fluctuations are computed using a standard approach \citep[e.g.,][]{pope2000book,chhiber2021ApJ_psp}: \(\tilde{\bm{v}} = \tilde{\bm{V}} - \bm{V}\) and \(\tilde{\bm{b}} = \tilde{\bm{B}} - \bm{B}\), where \(\bm{B}\equiv \langle\tilde{\bm{B}}\rangle\) is the mean magnetic field in an interval. The average energies per-unit-mass in magnetic and velocity fluctuations are then \(\langle \tilde{v}^2\rangle/2\) and \(\langle \tilde{b}^2\rangle/2\), respectively, and (twice) the total turbulent energy is \(Z^2=v^2 + b^2\), where we have defined \(v^2\equiv\langle \tilde{v}^2\rangle\) and \(b^2\equiv\langle \tilde{b}^2\rangle\) as the mean-squared fluctuations. We neglect density fluctuations since inner-heliospheric turbulence is believed to be nearly incompressible \citep{belcher1971JGR,matthaeus1990JGR,Zank2017ApJ835}. Note that our averaging procedure obtains information about the energy-containing scales of turbulence, at which the largest turbulent structures (or ``eddies'') inject energy into the inertial-range cascade \citep[e.g.,][]{pope2000book,Kiyani2015RSPTA,bandyopadhyay2020ApJS_cascade,Wu2022ApJ}.

\subsection{Probability distributions and the speed-turbulence correlation}

\begin{figure}
\centering
\includegraphics[width=.65\textwidth]{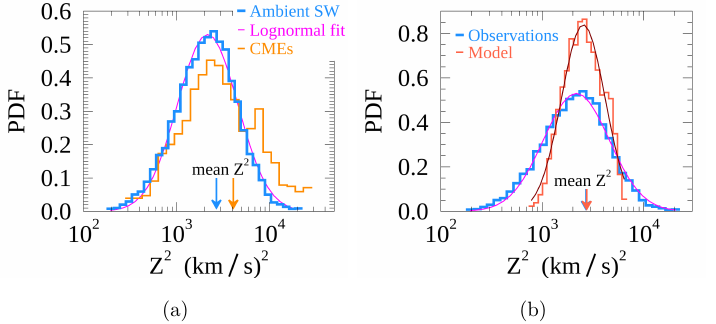}
\caption{Probability density functions (PDFs) of log of turbulence energy (\(\log Z^2\)), with horizontal axes showing corresponding \(Z^2\) values on a logarithmic scale. (a): Thick blue histogram is for ambient solar wind (sw) intervals and thin orange histogram is for intervals that include ICMEs. The best-fit Gaussian to \(\log Z^2\) (i.e., a log-normal fit to \(Z^2\)) for the ambient sw is shown as a magenta curve. Sample means are marked with arrows of corresponding color; with \(1\sigma\) spread, these are \(2.7\pm2.1 \times10^3 ~(\kmps)^2\) and \(4.1\pm4.0\times10^3~(\kmps)^2\), respectively. (b) The observed PDF for ambient sw is compared with the PDF obtained from the modeled \(Z^2\) (thin tomato-colored histogram) and its best-fit log-normal (maroon curve). The model shown here is the quadratic fit from Table \ref{tab:fit_paramas}. Sample means for the two cases are marked with arrows as in (a); these almost overlap. The model \(1\sigma\) is \(1.2\times10^3 ~(\kmps)^2\). Gaussian fits are performed using the IDL function {\fontfamily{cmtt}\selectfont
gaussfit.pro}. In all PDFs only bins with at least 10 counts are displayed.}
\label{fig:pdf}
\end{figure}

Fig. \ref{fig:pdf}(a) shows probability density functions (PDFs) of \(\log Z^2\) for the ambient and CME intervals. Although our focus remains on the former, we note that the CME distribution is shifted to larger \(Z^2\) and has a mean that is \(\sim40\%\) greater than that for the ambient intervals, which is qualitatively consistent with a recent study by \cite{Good2023ApJ}. Both PDFs have large variance and appear to follow log-normal distributions \citep[i.e., \(\log Z^2\) is normally distributed;][]{Limpert2001Biosci}, as is often the case with turbulence properties observed in the heliosphere \citep[e.g.,][]{Burlaga2000JGR,padhye2001JGR,Ruiz2014SoPh,Pradata2025ApJ,Chhiber2025MNRAS}. A best-fit log-normal to the blue PDF shows very good agreement. Fig. \ref{fig:pdf}(b) is discussed below.

\begin{figure}
\centering
\includegraphics[width=0.4\textwidth]{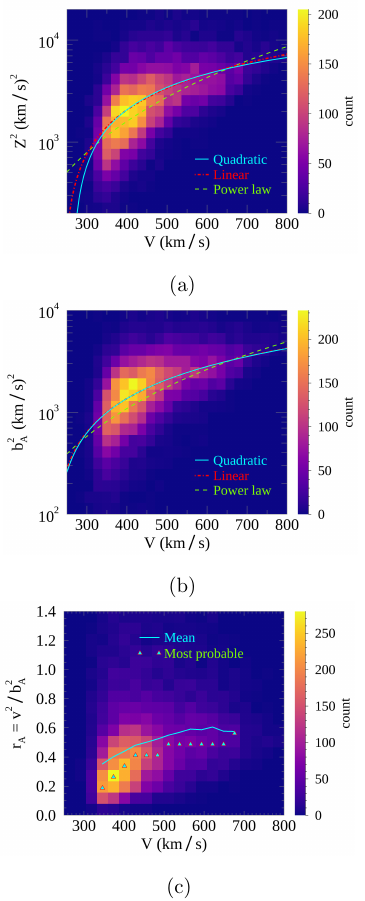} 
\caption{Joint histograms of mean speed \(V\) and (a) total turbulence energy \(Z^2\), (b) magnetic turbulence energy \(b^2_A\), and (c) Alfv\'en ratio \(r_A\). Colorbars show interval abundance. Panels (a) and (b) show three empirical fits each of \(Z^2\) and \(b^2_A\) to \(V\), respectively. Fit parameters are stated in Table \ref{tab:fit_paramas}. Panel (c) shows the mean and most probable values of \(r_A\) in bins of \(V\).}
\label{fig:2dhist}
\end{figure}

Fig. \ref{fig:2dhist} shows colormaps of two-dimensional (2D) joint histograms of \(V\) paired with three turbulent quantities. A clear positive correlation is evident between mean speed and the two turbulence energies, \(Z^2\) and \(b^2\). The Pearson correlation coefficients (PCCs) between \(V\) and \(\log Z^2\) and \(\log b^2\) are 0.59 and 0.56, respectively. We have similarly evaluated the relationship between the absolute value of the mean magnetic field and \(Z^2\), but the correlation is rather weak (\(\sim -0.04\)). Empirical fits to three different functions are shown, with fit parameters specified in Table \ref{tab:fit_paramas}. The quadratic and linear fits hardly differ in the core of the distribution, and the former has the lowest mean-squared error of the three (unsurprisingly, with its three free parameters). The ratio of velocity and magnetic fluctuation energies, called the Alfv\'en ratio (\(r_A = v^2/b^2\)), indicates relatively smaller variation compared with the turbulence energies; its mean changes from \(\sim 0.4\) in slow wind \( (V\lesssim500 ~\kmps) \) to \(\sim0.6\) at faster wind speeds, consistent with previous studies \citep{bruno2013LRSP}.

In addition to \(Z^2\), we have also shown fits of \(b^2\) to \(V\), and the joint distribution of \(r_A\) and \(V\),  since energetic-particle diffusion coefficients are usually expressed in terms of \(b^2\) \citep[e.g.,][]{Engelbrecht2022SSR}. In studies that apply solar wind models to CR/SEP transport, \(b^2\) is commonly obtained from \(Z^2\) by assuming a value for \(r_A\) \citep[e.g.,][]{guo2016ApJ,chhiber2017ApJS230}. Our suggested approach would directly obtain \(b^2\) from \(V\) by means of the quadratic fit in Table \ref{tab:fit_paramas}. The performance of this model is further evaluated in Fig. \ref{fig:time1}, below. Note that this approach will yield the magnetic fluctuation energy in Alfv\'en units, which can then be converted to magnetic-field units using (readily available) density data. Note also that we have performed a similar analysis of \textit{Wind} observations \citep{Wilson2021RevGeophs}, finding consistent results and further reinforcing their robustness.

\begin{table}[]
  \centering
  \vspace{0.25em}
  \begingroup
  \small
  \setlength{\tabcolsep}{4pt}      
  \renewcommand{\arraystretch}{1.05}
  \setlength{\arrayrulewidth}{0.6pt}
  \begin{tabular}{l | ccc | cc | cc}
       & \multicolumn{3}{|c|}{Quadratic: \(y=A_0+A_1V+A_2V^2\)} & \multicolumn{2}{|c|}{Linear: \(y=B_0+B_1V\)} & \multicolumn{2}{|c}{Power law: \(y=C_0 V^{C_1}\)} \\
      \(y\) & $A_{0}$ & $A_{1}$ & $A_{2}$ & $B_{0}$ & $B_{1}$ & $C_{0}$ & $C_{1}$ \\
      \hhline{-|---|--|--}
      $Z^{2}$ & $-4633 \pm 521$ & $19.3 \pm 2.1$ & $-0.006 \pm 0.002$
      & $-3126 \pm 111$ & $13.0 \pm 0.2$ & $0.0007 \pm 0.1872$ & $2.43 \pm 0.03$ \\
      \hhline{-|---|--|--}
      $b^{2}$ & $-1625 \pm 269$ & $7.6 \pm 1.1$ & $-0.0004 \pm 0.0011$ & $-1527 \pm 58$ & $7.2 \pm 0.1$ & $0.002 \pm 0.185$ & $2.17 \pm 0.03$ \\
    \end{tabular}%
  
  \endgroup
  \caption{Empirical fit parameters with \(1\sigma\) uncertainty estimates for $y=Z^2$ and $y=b^2$, as functions of \(V\). Fitting was based on the full 25-year dataset (cf. Fig. \ref{fig:time1}). Units of coefficients are consistent with \(y\) in \((\kmps)^2\) and \(V\) in \(\kmps\). Quadratic fitting was performed using the IDL function {\fontfamily{cmtt}\selectfont poly\_fit.pro}, and linear and power-law fits were obtained using the IDL function {\fontfamily{cmtt}\selectfont linfit.pro}. A cubic fit was also performed, but was indistinguishable from the quadratic one.}
  \label{tab:fit_paramas}
\end{table}

For a statistical comparison of the model-generated values with the observed \(Z^2\), we plot the PDFs of the two in Fig. \ref{fig:pdf}(b); model values are computed from the mean speed of each interval using the quadratic fit (Table \ref{tab:fit_paramas}). Consistent with the observations, the model produces log-normally distributed \(Z^2\), with a sample mean that is nearly identical to that of the observations. Modeled values have a notably smaller spread, with a standard deviation that is half of the observed value.

\subsection{Time-series comparisons of observed and modeled \(Z^2\)}

\begin{figure}
\centering
\includegraphics[width=1.\textwidth]{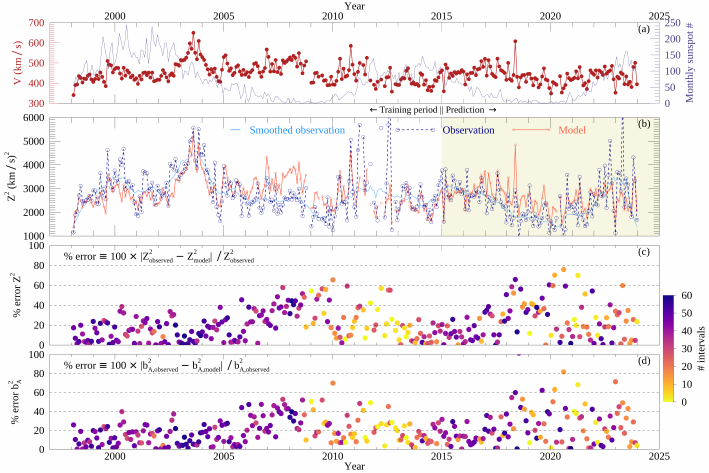}
\caption{Evaluation of model performance over a 25-year period.
{(a)}: For context, mean solar-wind speed is shown at 30-day cadence (see text) as brown connected circles that map to the left vertical axis. Monthly averages of solar sunspot number are shown as dark blue curve that maps to the right vertical axis.
{(b)}: Dashed dark-blue curve with circles shows observed \(Z^2\) at 30-day cadence; light-blue solid curve shows a 360-day running average. Red curve shows the modeled \(Z^2\) at 30-day cadence (see text for details). Here, model values are obtained from a quadratic fit - \(Z^2=A_0+A_1 V + A_2V^2\) - based on observations preceding 2015, i.e., from the annotated `Training period'; \(A_0=-3533,~A_1=15,~A_2=-0.002\) (cf. Table \ref{tab:fit_paramas}). Model values after the training period constitute a type of forecast, shown on a beige-shaded background. PCC between the 30-day cadence observations and model values is 0.61.
(c): Circles show percent error between observed and modeled values of \(Z^2\), at 30-day cadence. Color of circles maps to colorbar on the right, indicating number of 12-hour intervals within each 30-day period represented by a circle.
(d): As in (c), but for the quadratic fit for \(b_A^2\) instead of \(Z^2\), with \(A_0=-717,~A_1=4,~A_2=-0.003\).}
\label{fig:time1}
\end{figure}

The model's performance is further evaluated in Fig. \ref{fig:time1}. The top panel provides context, showing the mean solar-wind speed for the 25-year period; here the 12-hour cadence time-series of mean speed computed from ambient solar wind intervals is smoothed using a moving boxcar average spanning 30-days, and the result is plotted at a 30-day cadence. Also shown are monthly averages of the solar sunspot number obtained from \cite{SILSO_Sunspot_Number}. 

Panel (b) shows a comparison of the observed \(Z^2\) (smoothed and plotted at 30-day cadence, as above) with the quadratic model. To demonstrate the robustness of the model and to showcase its applicability to time periods beyond those in which the fitting was performed, for this figure we have computed a best-fit quadratic based on observations preceding 2015, defining a `training period'. Then the model values before 2015 are a `reconstruction', while those from 2015 onward constitute a type of prediction. Note that the parameters of this fit to the training period (see Fig. \ref{fig:time1} caption) are not significantly different from those obtained from the full 25-year dataset (Table \ref{tab:fit_paramas}). Note also that the model \(Z^2\) in Fig. \ref{fig:time1} is computed by applying the quadratic fit to the smoothed timeseries of \(V\) described above, and then plotted at 30-day cadence.

Overall, the model appears to give a good-to-reasonable agreement with observations in both the training and prediction periods. There are instances of remarkably detailed quantitative agreement for a range of observed \(Z^2\) (and speed), while a few instances indicate discrepancy by a factor of \(\sim 2\). The PCC between the 30-day cadenced observations and the model values is 0.61. The 360-day smoothed observation (light blue curve) indicates that the model follows the observed long-term trends. We note that some observational periods are highly ``noisy'', with large spikes in the dashed blue curve. This is likely related to a paucity of observations in those periods, as discussed below. We also note a moderate correlation between sunspot number and observed turbulence energy, with a PCC of \(\sim 0.4\). The corresponding time-series for \(b^2\) (not shown) is very similar to the \(Z^2\) case, including the level of visual agreement between the model and observations.

\begin{figure}
\centering
\includegraphics[width=.35\textwidth]{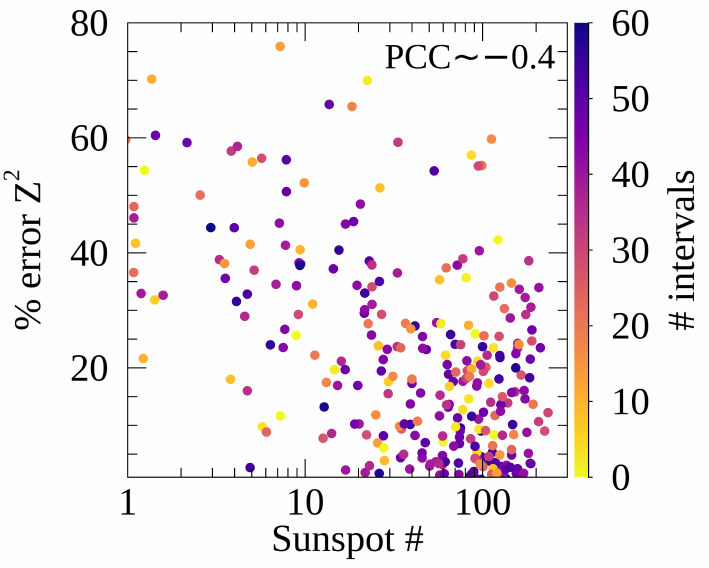}
\caption{Scatterplot of percent error between modeled and observed \(Z^2\) (at 30-day cadence) and monthly sunspot number (both from Fig. \ref{fig:time1}). The PCC is \(-0.37\). The colorbar indicates number of 12-hour intervals within each 30-day period represented by a circle.}
\label{fig:err_sunspot}
\end{figure}

Panel (c) of Fig. \ref{fig:time1} shows the percent error between the observed and modeled \(Z^2\) at a 30-day timescale; this is generally below 20\% for around half of the 25-year period, but there are periods when it becomes more variable, reaching 60\% on occasion. There is a moderate anti-correlation between sunspot number and the error (PCC \(-0.37\)), seen also in the scatter plot in Fig. \ref{fig:err_sunspot}. This suggests that the model may perform better during solar maximum when \textit{ACE} is more likely to sample coronal-hole wind \citep[e.g.,][]{mccomas2003grl}% This hypothesis is confounded by the simultaneous anti-correlation between percent error and the number of intervals (PCC \(\sim -0.15\))
; further study will be required on this point. We note that the model's better performance during solar maximum makes it suitable for application to SEP forecasting, since these events are more numerous during high solar activity. Panel (d) shows the corresponding error for \(b_A^2\). Again, it is very similar to the \(Z^2\) case, which is not surprising in view of their similar distributions as functions of \(V\), and the small variation in \(r_A\) (Fig. \ref{fig:2dhist}).

The color of the circles in panels (c) and (d) of Fig. \ref{fig:time1} and in Fig. \ref{fig:err_sunspot} indicates the number of 12-hour intervals that were available within each 30-day period represented by a circle. It is evident that the spiky values of observed \(Z^2\) in the middle panel tend to occur when there is a paucity of intervals (indicated by light-colored circles). The factors that produce the observed variations in the number of intervals were discussed in Sec. \ref{sec:data}. 

\begin{figure}
\centering
\includegraphics[width=.75\textwidth]{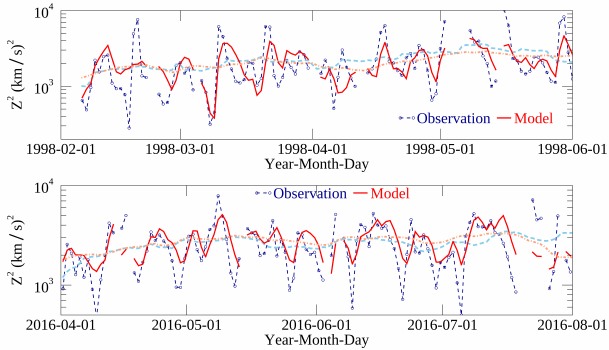}
\caption{A ``zoomed in'' comparison of observed and modeled \(Z^2\), spanning a period of around 4 months. Top and bottom panel cover periods that are in the ``training'' and ``prediction'' time ranges, respectively. Dashed dark-blue curve with circles shows observations smoothed and plotted at a 1-day cadence, with the corresponding model curve obtained from a daily-smoothed \(V\) and plotted in red at 1-day cadence. The model used is the quadratic fit for \(Z^2\) from Fig. \ref{fig:time1}'s caption. Pale blue curve with long dashes and dash-dotted red curve show the respective 10-day running averages. The PCC between the daily observations and model values is 0.61 and 0.55 in the top and bottom panels, respectively.}
\label{fig:time2}
\end{figure}

Fig. \ref{fig:time2} shows a ``zoomed in'' comparison of observed and modeled \(Z^2\), both evaluated at a daily cadence in this instance, using an approach analogous to the one used to produce the 30-day cadence values above. The top panel covers a time period within the ``training'' time range, while the bottom panel is from the ``prediction'' time range. Even at the much shorter timescales (compared to those in Fig. \ref{fig:time1}) we observe periods of remarkably good agreement, and the 10-day running averages overlap well. The PCC between observations and modeled values (at daily cadence) is 0.61 and 0.55 in the top and bottom panels, respectively. Similar results are obtained for \(b^2\) (not shown).

\subsection{Ensemble forecasts and probability distributions of errors}

\begin{figure}
\centering
\includegraphics[width=.65\textwidth]{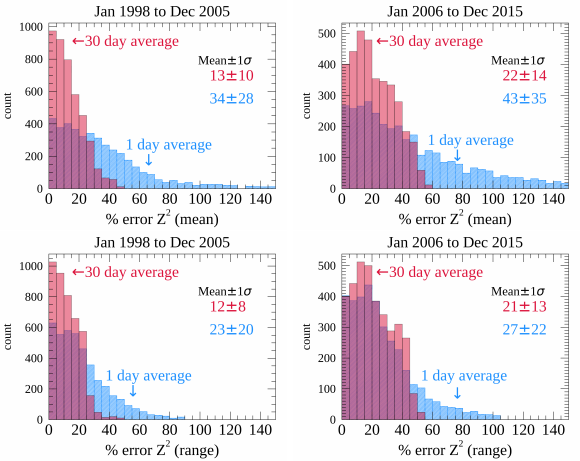}
\caption{Histograms of the percent error between observed and modeled \(Z^2\). Left and right panel columns correspond to two time periods. Top and bottom panel rows correspond to two methods of evaluating the \% error, labeled here as `mean' and `range' on the horizontal axes, with the latter based on an ensemble modeling approach (see text). Peach and blue (with diagonal stripes) histograms correspond to the error evaluated at timescales of 30 days and 1 day, respectively, as described in the text. The bin width in each histogram is 10\%, and only bins with count > 10 are shown. The mean\(\pm 1\sigma\) values of \% error for the 30 day and 1 day timescales are labeled in the respective colors in each panel.}
\label{fig:err}
\end{figure}

This section presents a more detailed error analysis. Fig. \ref{fig:err} shows histograms of the percent error between observed and modeled \(Z^2\) for two time periods: the left column panels cover Jan 1998 to Dec 2005, a period when the \% error in Fig. \ref{fig:time1} is relatively low, while the right column panels cover Jan 2006 to Dec 2015, when the \% error in Fig. \ref{fig:time1} is relatively large. Top and bottom rows of panels show the \% error evaluated using two different methods, labeled `mean' and `range' in the horizontal axes, where the latter is based on an ensemble modeling approach that accounts for uncertainty in the fit parameters to estimate a range of modeled \(Z^2\) values for each 12 hour interval. Within each panel, Fig. \ref{fig:err} displays histograms of the \% error computed at two timescales. A more detailed description follows below.

We first describe the top panel row of Fig. \ref{fig:err}. The peach colored histograms correspond to the error computed as \(\epsilon_\text{mean}\equiv100\times |Z^2_\text{obs,30} - Z^2_\text{mod,30}|/Z^2_\text{obs,30}\), where \(Z^2_\text{obs,30}\) is the observed \(Z^2\) at 12-hour cadence after being smoothed using a moving boxcar average of duration 30 days, and \(Z^2_\text{mod,30}\) is obtained by computing the quadratic fit (Table \ref{tab:fit_paramas}) on a 30-day smoothed \(V\). Note that the difference between this \% error and that shown in Fig. \ref{fig:time1} is that the latter is further down-sampled from a 12-hour cadence to a 30-day cadence. The blue histogram with stripes shows the analogous result when the moving boxcar average has a duration of 1 day. Examining the histograms and the associated mean\(\pm1\sigma\) values, we observe that the model is more accurate when used to obtain relatively long-term averages (in this case, monthly) of turbulence energy, compared to its accuracy at a daily scale. Similar results are obtained for \(b^2\) (not shown).

The bottom panel row of Fig. \ref{fig:err} is based on an ensemble-style prediction approach. The \% error is evaluated for three cases: in addition to the `mean' case described above, we estimate upper and lower bounds for \(Z^2_\text{mod}\) from the \(1\sigma\) uncertainty estimates given in Table \ref{tab:fit_paramas}, yielding \(Z^2_\text{mod,up}\) and \(Z^2_\text{mod,lo}\) respectively, where \(Z^2_\text{mod}\) is obtained from either the 1-day or the 30-day smoothed \(V\) (at 12-hour cadence) described above. To be more precise, the quadratic fit parameters for \(Z^2_\text{mod,up}\) are \(A_0 = -4633+521,~A_1=19.3+2.1,~A_2=-0.006+0.002\), and those for \(Z^2_\text{mod,lo}\) are \(A_0=-4633-521,~A_1=19.3-2.1,~A_2=-0.006-0.002\) (see Table \ref{tab:fit_paramas}). The associated \% error is then computed for the three cases: \(\epsilon_\text{mean}\) as described above, \(\epsilon_\text{lo}\equiv100\times |Z^2_\text{obs} - Z^2_\text{mod,lo}|/Z^2_\text{obs}\), and \(\epsilon_\text{up}\equiv100\times |Z^2_\text{obs} - Z^2_\text{mod,up}|/Z^2_\text{obs}\), where \(Z^2_\text{obs}\) is the observed \(Z^2\) at 12-hour cadence with 1-day or 30-day smoothing, matching the smoothing scale of \(Z^2_\text{mod}\). 

For each 12-hour interval we thus get three values of \% error for a particular smoothing scale (here, 1 or 30 day): \(\{\epsilon_\text{mean},\epsilon_\text{lo},\epsilon_\text{up}\}\). Note that ``mean'', ``lo'', and ``up'' here refer to the ``mean'' and lower/upper bounds (respectively) of \(Z^2_\text{mod}\) from which \(\epsilon\) is computed, and not to the magnitude of \(\epsilon\) itself. For each 12-hour interval we identify the minimum \(\epsilon\) from this set: \(\epsilon_\text{range}\equiv\text{min}~\{\epsilon_\text{mean},\epsilon_\text{lo},\epsilon_\text{up}\}\), which provides an error estimate of the best-performing realization within the ensemble of modeled \(Z^2\). This procedure is carried out for both 1 day and 30 day smoothing cases within the two selected time periods (Jan 1998 to Dec 2005 and Jan 2006 - Dec 2015), producing the bottom panels of Fig. \ref{fig:err}. Comparing the histograms and associated mean\(\pm1\sigma\) values between the top and bottom rows, it is evident that the ensemble approach yields a significant reduction in model error for the 1-day timescale, while the 30-day timescale remains essentially unaffected.

%\textbf{xx Add shaded region defined by uncertainty in model parameters }

\section{Conclusions}

In this paper we have investigated the relationship between bulk flow speed and MHD-scale turbulence energy in the near-Earth solar wind by analyzing 25 years of \textit{ACE} observations. The key result is an empirical law that produces reasonably accurate predictions of the average turbulence energy, simply from coarse-grained speed data. For the total and magnetic turbulence energies (\(Z^2\) and \(b^2\), respectively, where the latter is in Alfv\'en units) these relations are (Table \ref{tab:fit_paramas}):
\begin{itemize}
    \item \(Z^2=(-4633\pm521) + (19\pm2)~V + (-0.006\pm0.002)~V^2\)
    \item \(b^2 = (-1625\pm269) + (8\pm1)~V + (-0.0004\pm0.0011)~V^2\)
\end{itemize}
We suggest that the proposed approach (likely the first of its type, although similar in spirit to the widely-used WSA model) has the potential to enhance the capabilities of operational space-weather models and other datasets that are typically not considered suitable for turbulence studies. For example, SEP forecasting models can leverage their speed data to obtain self-consistent and data-constrained estimates of turbulence energy, which is a crucial parameter for modeling SEP diffusion \citep[e.g.,][]{Engelbrecht2022SSR,Hu2022SciAdv,Whitman2023AdSpR}. Our empirical model may also be useful in global MHD simulations of the heliosphere that lack turbulence modeling \cite[e.g.,][]{Provornikova2024ApJ,Samara2024ApJ}. Following further validation of the empirical law(s) at greater/smaller helioradii \citep{Shi2023ApJ} and outside the ecliptic plane \citep[e.g.,][]{mccomas2003grl}, they can be applied to global flow maps from remote-imaging instruments like \textit{PUNCH} \citep{DeForest2025arXiv_PUNCH,Attie2025AAS} to extract turbulence levels spanning the meridional plane \citep[e.g.,][]{Usmanov2025ApJ}. The observed positive correlation between speed and turbulence amplitude may also be relevant to geomagnetic activity, which is known to be driven by high-speed streams \citep{DAmicis2007GRL,Baker2018SSR}. 

A number of aspects of our investigation underscore the models' robustness. We have employed a large \textit{ACE} dataset that covers more than two solar activity cycles to derive the empirical laws. We have also demonstrated their applicability to time periods beyond those in which the fitting was performed by separating the dataset into a ``training+reconstruction'' period and a ``prediction'' period. The model's performance at different timescales has been validated, and it is found that an ensemble forecast approach (that exploits the uncertainty in the model parameters) yields a significant reduction in error for the (more challenging) short time-scale case. Although our datasets are obtained from \(\text{L}_1\), we see no reason why the results would not be applicable to any location in the ecliptic at 1 AU (including \(\text{L}_{4/5}\)). We add as a caveat that the \textit{ACE} dataset employed here includes periods when the plasma measurements may be of low quality; although we have taken care to discard intervals that may be affected by such issues, it will be worthwhile in future work to re-validate the models using other 1 AU datasets, including the new \textit{IMAP} and \textit{Aditya \({L}_1\)} missions \citep{McComas2025SSR_IMAP,Tripathi2023IAU_Symp}. We have conducted preliminary analysis of observations from the \textit{Wind} spacecraft, which reinforce our conclusions.

In addition to validation of the empirical model(s) in heliospheric regions \textit{beyond} 1 AU, future work can also investigate its apparent superior performance during solar maximum. On a broader note, the observed speed-turbulence correlation is likely to be related to the well-known correlation between speed and proton temperature \citep[e.g.,][]{Matthaeus2006JGR,Demoulin2009SoPh,Elliott2012JGR,Shi2023ApJ}; further study of these correlations can help improve our understanding of the role of turbulence in solar wind acceleration and heating.

\vspace{5mm}
\hrule
\section*{Data Sources}
\textit{ACE} magnetic field data were downloaded
from \url{https://spdf.gsfc.nasa.gov/pub/data/ace/mag/level 2 cdaweb/mfih3/} and plasma data were downloaded from \url{https://spdf.gsfc.nasa.gov/pub/data/ace/swepam/level2_hdf/ions_64sec}. The ICME list is available at \url{https://doi.org/10.7910/DVN/C2MHTH}. Sunspot-number data was obtained from the World Data Center SILSO, Royal Observatory of Belgium, Brussels, \url{https://doi.org/10.24414/qnza-ac80}.

\section*{Acknowledgements}
This research was supported by NASA under the Living With a Star (LWS) Science program grant 80NSSC22K1020, and utilized resources provided by the  National Energy Research Scientific Computing Center (NERSC).



%\bibliographystyle{plainnat}
%\bibliography{chhibref}

% Don't change these lines
%\bsp	% typesetting comment
%\label{lastpage}
\end{document}